\documentclass[lettersize,journal]{IEEEtran}
\usepackage{subcaption}

\usepackage{amsmath,amsfonts}
\usepackage{algorithm}
\usepackage{array}
\usepackage[caption=false,font=normalsize,labelfont=sf,textfont=sf]{subfig}
\usepackage{textcomp}
\usepackage{stfloats}
\usepackage{url}
\usepackage{verbatim}
\usepackage{graphicx}
\usepackage{cite}
\usepackage{url}
\usepackage{bm}
\usepackage{xcolor}
\usepackage{amssymb}
\usepackage{amsthm}
\usepackage{booktabs}
\usepackage{multirow}
\usepackage{threeparttable}
\usepackage{algpseudocode}

\newtheorem{prop}{Proposition}
\newtheorem*{remark}{Remark}
\begin{document}

\title{Structure Before Strength: Physics-Informed Reference Eikonal Learning for High-Frequency Radio Map Construction}

\author{Sicong Ye,~\IEEEmembership{Graduate Student Member,~IEEE,} Sibo Zhang, Yichen Luo, Ming Xiao,~\IEEEmembership{Senior Member,~IEEE}, and H. Vincent Poor~\IEEEmembership{Life Fellow,~IEEE}
\thanks{Sicong Ye, Sibo Zhang, Yichen Luo, and Ming Xiao are with the Department of Information Science and Engineering, KTH Royal Institute of Technology, 10044 Stockholm, Sweden (e-mail: sicongy@kth.se; zhang.sibo@outlook.com; yichenlu@kth.se; mingx@kth.se).
H. Vincent Poor is with the Department of Electrical and Computer Engineering, Princeton University, Princeton, NJ 08544 USA (e-mail: poor@princeton.edu).}
\thanks{This work has been submitted to the IEEE for possible publication. Copyright may be transferred without notice, after which this version may no longer be accessible.}
}

\markboth{Journal of \LaTeX\ Class Files,~Vol.~14, No.~8, August~2021}%
{Shell \MakeLowercase{\textit{et al.}}: A Sample Article Using IEEEtran.cls for IEEE Journals}


 \maketitle

\begin{abstract}
Accurate radio maps provide spatial channel knowledge for environment-aware wireless communications. However, their construction remains challenging in high-frequency multipath environments, where the underlying wave field exhibits rapid wavelength-scale spatial oscillations and complicated multipath interference. 
This paper proposes a propagation-structure-aware framework for received signal strength (RSS) radio-map construction.
Based on high-frequency asymptotic analysis, we introduce a reference eikonal function that yields a direction-dependent redistribution of local spatial frequencies. We quantify the resulting branch-wise spatial frequency in terms of directional mismatch with the reference direction and identify an angular condition for spatial-frequency suppression.
We further construct a computationally tractable surrogate reference eikonal that approximates the dominant propagation structure, and characterize its piecewise behavior in terms of smooth regions and critical regions.
Building on this characterization, we develop a two-stage learning framework that first employs a physics-informed neural network (PINN) to estimate the reference eikonal under joint physics and data supervision, and subsequently exploits it as propagation-aware auxiliary information for RSS radio-map prediction.
Experiments demonstrate that the proposed framework consistently achieves the lowest normalized mean squared error (NMSE) across diverse propagation settings, with consistent improvements across different backbones and enhanced training-data efficiency.
The reference-eikonal construction is computationally efficient and yields accurate estimates that remain robust to moderate reference-direction mismatch, while ablation studies confirm the effectiveness of the proposed method.
\end{abstract}

\begin{IEEEkeywords}
Radio map, eikonal function, high-frequency asymptotics, physics-informed neural network, multipath propagation, received signal strength.
\end{IEEEkeywords}

\section{Introduction}
\IEEEPARstart{E}{merging} sixth-generation (6G) wireless networks are expected to evolve toward increasingly environment-aware systems \cite{10430216,8766143,10734395}, where the spatial knowledge of the surrounding propagation environment is exploited to support adaptive and intelligent transmission. 
This paradigm calls for extending channel knowledge beyond link-specific instantaneous channel state information (CSI) toward site-specific characterization of spatial propagation conditions.
Such site-specific propagation knowledge can be organized into a radio map, which characterizes the spatial distribution of location-dependent wireless quantities, such as received signal strength (RSS), over the geographical region of interest\cite{9931518,11677060}. 
Radio maps can support a wide range of applications \cite{9931518,11677060,11433407}, including network planning, coverage optimization, mobility management, resource allocation, localization, and proactive transmission control. 
Despite their broad applications, constructing accurate radio maps efficiently remains challenging, particularly with limited or no spatial measurements and in complex and heterogeneous propagation environments.

\subsection{Related Work}

Conventional radio-map construction approaches span a broad spectrum of trade-offs between computational efficiency and propagation fidelity.
Simplified analytical and empirical models, together with spatial interpolation and reconstruction techniques such as kriging \cite{7817747} and matrix completion \cite{9992123}, provide low computational complexity for radio-map construction. However, these methods often struggle to capture site-specific propagation characteristics and spatial variations in complex and heterogeneous environments. 
For higher-fidelity propagation modeling, physics-based methods, including full-wave electromagnetic solvers and ray tracing \cite{10465179,9459462,580789}, explicitly account for wave interactions with the surrounding environment.
While providing detailed site-specific predictions, these methods require accurate geometric and material information and can incur substantial computational costs, particularly in electrically large and geometrically complex environments.
Alternatively, dense measurements of the wireless field provide direct, site-specific propagation information but incur substantial measurement, deployment, and maintenance costs, especially in large-scale and dynamic environments.
Overall, conventional approaches face persistent trade-offs among propagation fidelity, computational complexity, and measurement overhead, limiting their scalability and adaptability to diverse and rapidly changing wireless environments envisioned for 6G networks.

To overcome these limitations, end-to-end learning-based approaches have been extensively investigated for radio-map construction, directly learning the mapping from environmental information and available observations to spatial radio distributions. By shifting much of the computational burden to offline training, these approaches enable efficient radio-map construction through network inference.
A representative example is RadioUNet \cite{9354041}, which formulates radio-map estimation as an image-to-image learning problem and employs convolutional neural networks (CNNs) to exploit the spatial structure of environmental layouts and propagation measurements for fast radio-map inference.
Furthermore, more advanced neural network architectures, such as graph neural networks (GNNs) \cite{10078269,li2024radiogat} and Kolmogorov–Arnold networks (KANs) \cite{11516172}, have also been investigated for radio map construction to exploit spatial relational structures and enhance nonlinear function approximation, respectively.
More recently, a hybrid Mamba–UNet architecture has been developed to jointly capture long-range spatial dependencies and local propagation features, improving radio-map construction accuracy while maintaining computational efficiency \cite{11190042}.
Generative models, including generative adversarial networks (GANs) and diffusion models, have also been explored to capture complex spatial patterns in radio maps. Specifically, conditional GANs leverage adversarial learning to reconstruct radio maps from sparse measurements \cite{10130091}, while diffusion-based models employ iterative denoising for sampling-free radio-map construction conditioned on environmental information \cite{10764739}.
Despite advances in network architectures and generative modeling, most data-driven approaches learn statistical relationships between environmental inputs and radio maps without explicitly incorporating the physical mechanisms and spatial structures governing radio-wave propagation. Consequently, their prediction accuracy and generalization remain sensitive to the availability and representativeness of training data, particularly under limited training data or distribution shifts across propagation environments \cite{10352759}. 
These limitations have motivated the integration of propagation knowledge into the learning frameworks to improve data efficiency, physical consistency, interpretability and generalization. 


Recent studies have increasingly explored the integration of radio-propagation knowledge into learning-based radio map construction \cite{10634040}.
One important direction incorporates propagation physics directly into the learning objective through physics-informed neural networks (PINNs) \cite{10608081}. 
By introducing physics-based loss functions derived from governing electromagnetic equations, these approaches encourage the learned fields to conform to the underlying propagation laws.
Such physics-constrained learning can improve physical consistency and reduce dependence on densely labeled data. Its application to high-frequency radio-map construction, however, is substantially more challenging because the field contains rapidly oscillatory components and may exhibit non-smooth behaviors near critical propagation regions and material interfaces. Directly constraining such a field therefore requires the learning model to simultaneously resolve fine-scale oscillations and satisfy local differential relations, potentially complicating optimization.
Another class of approaches introduces propagation knowledge through auxiliary propagation maps supplied to the radio-map prediction model \cite{10130091,11516172,Guo2026RMPrior,li2024radiogat}. These maps can be constructed from analytical propagation models, such as path-loss-based models fitted to sparse observations, or learned from measurements or simulated data to provide coarse coverage estimates and propagation-aware  representations. By providing the downstream radio-map predictor with large-scale propagation structure, these auxiliary maps complement raw environmental and transmitter information with physically informative guidance.
Their effectiveness, however, critically depends on the fidelity of the constructed representations, which can be difficult to maintain in complex propagation environments. Analytical representations may suffer from model mismatch due to simplified or inaccurate propagation assumptions, whereas data-driven representations may lack physical interpretability and fail to preserve propagation consistency in the absence of appropriate physical constraints.
Additionally, propagation-mechanism-aware representations have been introduced by explicitly identifying structures associated with specific propagation phenomena, such as shadowing regions and visibility information \cite{11587950,11278649,10424680}. While these handcrafted features can improve the construction of propagation-specific spatial features, their reliance on heuristic rules limits their ability to comprehensively characterize complex propagation phenomena.
Moreover, increasing the fidelity or completeness of propagation information does not necessarily translate into commensurate improvements in radio-map construction \cite{jaensch2026radio}, despite the additional measurement, environmental characterization, and computational costs required to obtain more detailed physical information.

\subsection{Motivation and Contributions}
These observations suggest that the key challenge lies not merely in increasing model capacity or incorporating more detailed propagation information, but in identifying a compact physical representation that captures the source of spatial complexity in radio-map construction.
To this end, we revisit radio propagation from a high-frequency asymptotic perspective, which reveals the local propagation directions as a basis for designing a structured propagation representation.
Constructing such a representation requires both an efficient means of obtaining the desired propagation information and dedicated strategies for handling distinct propagation characteristics across spatial regions.
These considerations motivate the proposed propagation-structure-aware radio-map construction framework. The main contributions are as follows.
\begin{itemize}
    \item \textbf{High-frequency propagation analysis and reference-eikonal formulation:}
    We provide a high-frequency asymptotic characterization of radio-map representation complexity and formulate a reference-eikonal transformation that redistributes branch-wise local spatial frequencies according to their directional alignment with a reference direction.
    We further derive the dependence of the residual spatial frequency on directional mismatch and establish an explicit angular criterion for reducing branch-wise spatial oscillations.
    \item \textbf{Reference eikonal construction and regional characterization:} 
    We construct a computationally efficient surrogate reference eikonal based on propagation-length and event-penalty costs that captures the dominant propagation geometry.
    We further characterize its piecewise structure in terms of smooth regions satisfying the classical eikonal relation and critical interfaces caused by branch switching and admissibility transitions.
    \item \textbf{Propagation-structure-aware radio map construction:}
    We develop a two-stage framework that first estimates a reference eikonal from environmental geometry and transmitter information and subsequently incorporates it as propagation-aware auxiliary information for RSS radio-map construction.
    Specifically, the first stage employs a PINN with a dual-head architecture to separate structural prediction from residual correction, together with a scaffold training strategy that anchors the reference field at critical regions to facilitate local eikonal consistency in smooth regions.
    \item \textbf{Comprehensive experimental validation:} 
    Experiments confirm the structural fidelity of the proposed reference-eikonal representation and its computational efficiency relative to high-fidelity ray tracing.
    The proposed framework consistently outperforms the evaluated comparison methods in normalized mean squared error (NMSE) across diverse propagation settings, while its consistent gains across network backbones and training-data scales further demonstrate the generality and effectiveness of the proposed representation. 
\end{itemize}

\section{System Model and Problem Formulation}
We first describe the considered radio propagation scenario, followed by the radio-map construction problem formulation. 
Consider a radio propagation environment defined over a spatial domain of valid receiver locations $\Omega \subset \mathbb{R}^d$, characterized by environmental information $\mathcal{M}$ and source information $\mathcal{X}$. Specifically, $\mathcal{M}$ describes the geometry and, when available, the propagation-related attributes of the surrounding objects, while $\mathcal{X}$ characterizes the transmitter and its associated excitation parameters.
For each receiver location $\bm{x}\in\Omega$, the RSS defines a spatial radio map
\begin{equation}
\mathcal{R}: \Omega \rightarrow \mathbb{R},
\end{equation}
where $\mathcal{R}(\bm{x})$ denotes the RSS at location $\bm{x}$. 
Given the environmental representation $\mathcal{M}$ and source information $\mathcal{X}$, a conventional data-driven approach seeks to directly approximate the underlying environment-to-RSS mapping as
\begin{equation}\label{eqn:mapping}
\hat{\mathcal{R}}
=
\mathcal{F}_{\bm{\psi}}
\left(
\mathcal{M},\mathcal{X}
\right),
\end{equation}
where $\bm{\psi}$ denotes the trainable model parameters.
The radio-map construction problem can therefore be formulated as
\begin{equation}
\min_{\bm{\psi}}
\mathcal{L}
\left(
\hat{\mathcal{R}},
\mathcal{R}
\right)
,
\end{equation}
where $\mathcal{L}(\cdot,\cdot)$ measures the difference between the constructed and ground-truth RSS maps.

Improving the approximation capability of $\mathcal{F}_{\bm{\psi}}$ through larger or more sophisticated architectures, or incorporating more detailed propagation information, may enhance prediction accuracy.
However, neither approach necessarily makes the underlying propagation structure explicitly accessible to the learning model, leaving the intrinsic complexity of learning the environment-to-RSS mapping largely unresolved.
This observation motivates the introduction of an intermediate physical representation that explicitly captures the propagation structure implicit in the environmental and source information. Specifically, we define
\begin{equation}
    \mathcal{S} = \mathcal{G} \left( \mathcal{M},\mathcal{X} \right),
\end{equation}
where $\mathcal{G}(\cdot)$ denotes the underlying physical mapping from $(\mathcal{M},\mathcal{X})$ to an intermediate representation $\mathcal{S}$. Rather than introducing additional information, $\mathcal{S}$ provides a physically structured representation of propagation characteristics that may be difficult for a finite-capacity learning model to extract directly from the original inputs.
Since $\mathcal{S}$ is fully determined by $(\mathcal{M},\mathcal{X})$, it introduces no additional information, but instead reorganizes the available information into a physics-informed representation.
The radio-map construction is then given by
\begin{equation}
    \hat{\mathcal{R}}
    =
    \mathcal{F}_{\bm{\phi}}
    \left(
    \mathcal{M},\mathcal{X}, \mathcal{S}
    \right),
\end{equation}
where $\bm{\phi}$ denotes the trainable parameters. 
Our hypothesis is that an appropriate $\mathcal{S}$ can transform the original propagation mapping into a representation that reduces its effective complexity, thereby facilitating radio-map construction under finite-capacity models.
The key question is therefore which physical characteristics of high-frequency propagation primarily contribute to the representation complexity of radio-map construction and how such characteristics can be represented in a compact and computationally tractable form.

\section{High-Frequency Propagation Analysis and Reference-Eikonal Representation}\label{sec:III}

\subsection{High-Frequency Asymptotic Propagation Model}

We first establish the high-frequency asymptotic propagation model underlying the propagation-structure representation developed in this work.
Consider a time-harmonic electromagnetic field over a spatial domain $\Omega$. Under the scalar-wave approximation, the complex-valued spatial field $u(\bm{x})\in\mathbb{C}$ in a source-free region satisfies the Helmholtz equation
\begin{equation}\label{eqn:Helmholtz}
    \nabla^2 u(\bm{x}) + k^2 n^2(\bm{x})u(\bm{x}) = 0,
    \qquad \bm{x}\in \Omega\setminus \left(\partial\Omega\cup\{\bm{x}_{\rm tx}\}\right), 
\end{equation}
where $k=2\pi/\lambda$ is the free-space wavenumber, $\lambda$ is the wavelength, $n(\bm{x})$ is the spatially varying refractive index, and $\bm x_\mathrm{tx}$ denotes the source location.
The Helmholtz equation in \eqref{eqn:Helmholtz} characterizes the underlying wave propagation. However, directly resolving $u(\bm x)$ becomes increasingly demanding as $k$ increases, since its spatial oscillation occurs on the wavelength scale.
For high-frequency propagation, i.e., when the characteristic environmental scale is much larger than the wavelength, the field can be represented through high-frequency asymptotic methods, such as Wentzel–Kramers–Brillouin (WKB) decomposition \cite{Bender1999Advanced}. At a given spatial point, the field is represented as a superposition of multiple WKB branches, each associated with a distinct propagation path. Consequently, with $\mathcal A(\bm x)$ denoting the set of geometrically admissible propagation branches connecting the transmitter $\bm x_{\mathrm{tx}}$ to $\bm x$, the wave field $u(\bm x)$ can be represented as
\begin{equation}\label{eqn:WKB}
    u(\bm x) = \sum_{\ell \in \mathcal{A}(\bm x)} u_\ell (\bm x) =
    \sum_{\ell \in \mathcal{A}(\bm x)} A_\ell (\bm x) e^{ j k S_\ell(\bm x) },
\end{equation}
where $A_\ell(\bm x)$ is a slowly varying complex amplitude associated with the $\ell$-th admissible propagation branch and $S_\ell(\bm x)$ is the corresponding eikonal function.
In contrast to the rapidly oscillating field $u_\ell(\bm x)$, both $A_\ell (\bm x)$ and $S_\ell(\bm x)$ vary on the environmental scale and are asymptotically independent of the carrier frequency. 
Substituting $u_\ell(\bm x)$ into \eqref{eqn:Helmholtz} yields
\begin{equation}
\begin{split}
    \Bigl[ A_\ell \left( n^2 - \left| \nabla S_\ell \right|^2 \right) k^2  + j \left( 2 \nabla A_\ell \cdot \nabla S_\ell  + A_\ell \nabla^2 S_\ell  \right)  k \\ + \nabla^2 A_\ell
     \Bigr]  e^{ j k S_\ell } = 0.
\end{split}
\end{equation} 
Under the high-frequency condition, the leading-order term $\mathcal{O}(k^2)$ yields the eikonal equation, given by
\begin{equation}\label{eqn:eikonal}
    \left| \nabla S_\ell (\bm x)\right|^2 = n^2(\bm x).
\end{equation}
The next-order term $\mathcal{O}(k)$ gives
\begin{equation}\label{eqn:transport}
    2 \nabla A_\ell (\bm x) \cdot \nabla S_\ell (\bm x) + A_\ell (\bm x) \nabla^2 S_\ell (\bm x) = 0,
\end{equation}
which is the corresponding transport equation describing the amplitude evolution along the propagation branch.
Accordingly, the high-frequency Helmholtz equation separates naturally into two coupled physical components. 
The eikonal equation determines the spatial propagation geometry, whereas the transport equation describes the amplitude variation along that geometry.

\subsection{Reference Eikonal Transformation and Direction-Selective Spatial-Frequency Redistribution}
The high-frequency asymptotic propagation model avoids directly solving the full-wave Helmholtz equation by instead solving the eikonal and transport equations. Nevertheless, the constructed wave field remains highly oscillatory in space due to the rapidly varying phase term. 
Specifically, within a region where the $\ell$-th branch is locally smooth, the spatial gradient of the corresponding field is given by
\begin{equation}\label{eqn:u_l'}
    \nabla u_\ell (\bm x) = e^{j k S_\ell (\bm x)} \left( \nabla A_\ell (\bm x) +  j k A_\ell (\bm x) \nabla S_\ell (\bm x) \right).
\end{equation}
Under the WKB approximation, $A_\ell (\bm x)$ and $S_\ell (\bm x)$, together with their spatial derivatives, vary on an $\mathcal{O}(1)$ spatial scale independent of the wavenumber $k$. Consequently, in the high-frequency limit $k \rightarrow \infty$, the phase gradient term dominates in \eqref{eqn:u_l'}, yielding the leading-order approximation
\begin{equation}
    \nabla u_\ell (\bm x) \sim
    j k A_\ell (\bm x) \nabla S_\ell (\bm x) e^{j k S_\ell (\bm x)}.
\end{equation}
Hence the normalized spatial variation satisfies
\begin{equation} \label{eqn:kappa}
    \frac{\| \nabla u_\ell (\bm x) \|}{|u_\ell (\bm x)|} \sim k\|\nabla S_\ell (\bm x)\| = \| \bm \kappa_\ell (\bm x) \|,
\end{equation}
where $\bm  \kappa_\ell (\bm x) $ denotes the local spatial wavevector characterizing the local spatial wavenumber of the $\ell$-th branch. 
More generally, for any fixed derivative order $m$, the magnitude of the $m$-th order spatial derivative scales asymptotically as
\begin{equation}\label{eqn:u_l'm}
    \| \nabla^m u_\ell (\bm x) \| \sim  |A_\ell (\bm x)| \| \bm \kappa_\ell (\bm x) \|^m.
\end{equation}

The relations in \eqref{eqn:kappa} and \eqref{eqn:u_l'm} reveal two complementary manifestations of the intrinsic complexity of a high-frequency wave field. First, $\| \bm \kappa_\ell\|$ in \eqref{eqn:kappa} determines its local oscillation scale, with larger $\| \bm \kappa_\ell\|$ corresponding to shorter local wavelengths and hence increasingly rapid spatial variations that must be resolved by a direct representation \cite{pmlr-v97-rahaman19a}. 
Second, the magnitude of the $m$-th- order spatial derivatives grows as $\| \bm \kappa_\ell \|^m$ in \eqref{eqn:u_l'm}, resulting in higher-order Sobolev seminorms as the carrier frequency increases and thereby making direct representation more challenging \cite{YAROTSKY2017103,doi:10.1142/S0219530519410021}.
Therefore, even when the amplitude and eikonal remain spatially smooth, direct representation of high-frequency wave fields becomes progressively more challenging as the frequency increases.

To reorganize the branch-dependent spatial-frequency structure of the high-frequency field, we introduce a reference eikonal function $S_{\mathrm{ref}} (\bm x)$ to analytically factor out a reference spatial phase component. 
Specifically, the total field can be expressed as
\begin{equation}
    u (\bm x) = \tilde{u} (\bm x) e^{j k S_{\mathrm{ref}} (\bm x)},
\end{equation}
where 
\begin{equation}\label{eqn:u_tilde}
\begin{split}
    \tilde{u} (\bm x) &= u (\bm x) e^{-j k S_{\mathrm{ref}} (\bm x)} \\
    &= \sum_{\ell \in \mathcal{A}(\bm x)} A_\ell (\bm x) e^{ j k \left(S_\ell (\bm x) - S_{\mathrm{ref}} (\bm x)\right) },
\end{split}
\end{equation}
denotes the reference eikonal transformed field. 
The transformation replaces the absolute eikonal phase function $S_\ell (\bm x)$ of each branch by its phase relative to a common propagation reference $S_\ell (\bm x) - S_{\mathrm{ref}} (\bm x)$. 
Additionally, since the reference eikonal transformation factor has unit modulus, the transformation preserves the field RSS, i.e.,
\begin{equation}
     | u (\bm x) |^2 = |\tilde{u} (\bm x) e^{j k S_{\mathrm{ref}} (\bm x)}|^2 = |\tilde{u} (\bm x) |^2. 
\end{equation}
Therefore, the reference eikonal function $S_{\mathrm{ref}}$ changes the representation of the propagation field without modifying the received-power quantity of interest.
More importantly, the transformation alters the local spatial frequency of each propagation branch. In particular, the spatial wavevector of the transformed wave field $\tilde{u}_\ell (\bm x)$ is given by
\begin{equation}\label{eqn:tilde_kappa}
    \tilde{\bm \kappa}_\ell (\bm x) = k \left( \nabla S_\ell (\bm x) - \nabla S_{\mathrm{ref}} (\bm x) \right). 
\end{equation}
Therefore, although the transformation is performed in the spatial domain, its effect on the oscillatory structure is inherently direction dependent through the relative orientation between $\nabla S_\ell (\bm x)$ and $\nabla S_{\mathrm{ref}}$.
The following proposition quantifies this directional dependence and establishes the angular condition under which the spatial oscillation is suppressed.

\begin{prop}\label{prop1}
    Consider a locally smooth propagation branch eikonal function $S_\ell(\bm x)$ and a reference eikonal function $S_{\mathrm{ref}}(\bm x)$ satisfying the same eikonal equation in \eqref{eqn:eikonal} and let $\theta_\ell(\bm x)\in [0,\pi]$ denote the angle between $\nabla S_\ell(\bm x)$ and $\nabla S_{\mathrm{ref}}(\bm{x})$. Under the reference eikonal transformation in \eqref{eqn:u_tilde}, the local spatial wavenumber of the $\ell$-th branch is scaled according to
    \begin{equation}\label{eqn:gamma}
        \gamma_\ell(\bm x) = 
        \frac{ \| \tilde{\bm \kappa}_\ell(\bm x) \|}{ \| \bm \kappa_\ell(\bm x) \|} = 
        2\sin \left( \frac{\theta_\ell(\bm x)}{2} \right).
    \end{equation}
    Moreover, for any fixed derivative order $m\ge 1$, the leading-order spatial derivative of the transformed branch satisfies
    \begin{equation}\label{eqn:tilde_u_l'm}
        \| \nabla^m \tilde{u}_\ell (\bm x) \| \sim  |A_\ell (\bm x)|   \| \tilde{\bm \kappa}_\ell  (\bm x) \|^m.
    \end{equation}
    When $\theta_\ell=0$, the carrier-scale term vanishes, leaving only the slowly varying amplitude.
    Consequently, the reference eikonal transformation suppresses the spatial oscillation of the $\ell$-th branch if and only if
    \begin{equation}\label{eqn:theta}
        \theta_\ell(\bm x)<\frac{\pi}{3}.
    \end{equation}
    In two dimensions, this condition defines a $\pm60^\circ$ angular sector centered on the reference eikonal propagation direction.
\end{prop}

\begin{proof}
    A detailed proof is provided in the Appendix. 
\end{proof}

\begin{figure}[t]
    \centering
    \includegraphics[width=0.58\linewidth]{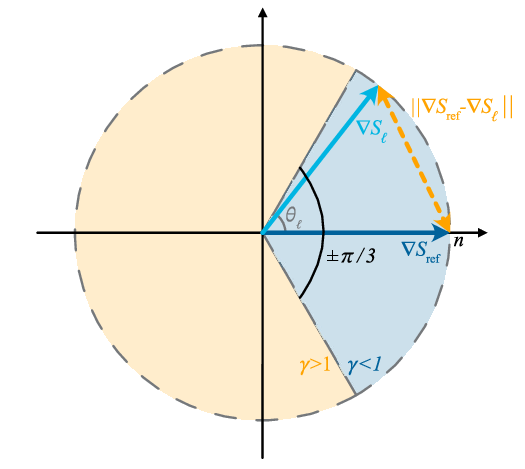}
    \caption{Geometrical interpretation of reference eikonal spatial transformation in the local gradient space.}
    \label{fig:pie}
\end{figure}

Proposition~\ref{prop1} reveals that the reference eikonal transformation produces a direction-selective regularization of the high-frequency spatial variation and admits a geometric interpretation, as illustrated in Fig.~\ref{fig:pie}.
At any fixed spatial location, the endpoints of the gradient vectors of all branch eikonal functions satisfying the same eikonal equation, i.e., $\|\nabla S_\ell\|=\|\nabla S_{\mathrm{ref}}\|=n$, lie on a eikonal circle of radius $n$ in the two-dimensional space. Accordingly, the transformed spatial wavevector in \eqref{eqn:tilde_kappa} can be geometrically represented by a chord connecting the endpoints of $\nabla S_{\mathrm{ref}}$ and $\nabla S_\ell$. 
Spatial frequency suppression therefore occurs precisely when the chord is shorter than the radius of the eikonal circle, which is equivalent to the $\pm60^\circ$ angular sector. In contrast, for branches outside this angular sector, the chord exceeds the circle radius, and the transformed field becomes increasingly oscillatory as the angular deviation from the reference direction increases.
Therefore, the proposed reference eikonal transformation does not universally reduce field oscillations, but rather redistributes the spatial frequency around the reference direction. 

\subsection{Reference Eikonal Construction}
Proposition~\ref{prop1} establishes that reference eikonal transformation yields a direction-selective redistribution of the local spatial frequencies. Although any eikonal function satisfying the same eikonal equation \eqref{eqn:eikonal} can serve as an admissible reference, its gradient direction determines which propagation branches benefit from spatial-frequency suppression. Therefore, an appropriate construction of $S_{\mathrm{ref}}$ is essential.

Since different propagation branches contribute unequally to the received field, we select the strongest propagation branch as reference, i.e., 
\begin{equation}
    \ell_{\mathrm{ref}}(\bm x)
    =
    \arg\max_{\ell}
    |A_\ell(\bm x)|^2. 
\end{equation}
Thus, the total transformed field in \eqref{eqn:u_tilde} can be rewritten as
\begin{equation}
    \tilde{u} = A_{\mathrm{ref}} + \sum_{\ell \neq \ell_{\mathrm{ref}}} A_\ell e^{ j k \left(S_\ell - S_{\mathrm{ref}} \right) },
\end{equation}
where $|A_{\mathrm{ref}} |= |A_{\ell_{\mathrm{ref}}}| = \max_{\ell} |A_\ell|$ and $S_{\mathrm{ref}}=S_{\ell_{\mathrm{ref}}}$. As a result, the carrier-scale oscillation of the strongest branch is eliminated completely, leaving the corresponding slowly varying amplitude.
Additionally, the selection of the strongest propagation branch also benefits the representation of the remaining branches when the received power is angularly concentrated around the dominant arrival direction.
According to Proposition~\ref{prop1}, every branch included in the angular sector \eqref{eqn:theta} satisfies $\gamma_\ell<1$, and therefore exhibits a lower residual spatial frequency after the transformation.

\begin{remark}
    Multi-frequency channel measurements \cite{11160744,11161884,10978475,9328091} and standardized channel models \cite{3GPP_TR38901_Rel19} have reported a tendency toward smaller omnidirectional azimuth spreads of arrival (ASA) at higher carrier frequencies in various indoor and outdoor representative scenarios. 
    This trend does not necessarily imply that all propagation branches lie within the $\pm 60^\circ$ angular sector. Nevertheless, it suggests that a larger proportion of the power-significant branches is likely to be confined within this sector.
    Thus, selecting the strongest power branch provides a physically interpretable reference that not only eliminates the oscillation of the dominant component but also reduces the residual spatial frequencies of angularly concentrated propagation branches with significant power.
\end{remark}

The reference eikonal defined above requires the branch-wise amplitudes and eikonal functions, which are generally unavailable from sparse radio measurements or environmental layouts. 
Nevertheless, Proposition~\ref{prop1} shows that spatial-variation suppression depends on the reference-eikonal gradient rather than its absolute value. This motivates the construction of a computationally accessible surrogate that approximates the dominant propagation branch while preserving its local propagation direction.
For the $\ell$-th branch, define its accumulated eikonal length as
\begin{equation}\label{eqn:L}
    L_\ell(\bm x) = \int_{\Gamma_\ell (\bm x)} n(\bm x')\, \| \mathrm{d} x' \|,
\end{equation}
where $\Gamma_\ell(\bm x)$ denotes the corresponding propagation trajectory satisfying the local eikonal condition. In the homogeneous free-space regions, \eqref{eqn:L} reduces to the accumulated physical propagation distance.
To construct a computable reference eikonal associated with the strongest branch, the following branch selection principles are adopted. First, shorter propagation paths generally experience lower attenuation in a homogeneous medium and are therefore prioritized. Second, different propagation mechanisms, such as reflection and diffraction, introduce additional boundary-dependent attenuation that cannot be characterized by path length alone.
Therefore, motivated by these considerations, we define propagation mechanism-aware cost of the $\ell$-th branch as
\begin{equation}\label{eqn:cost}
    J_\ell(\bm x) = L_\ell(\bm x) + \sum_{e \in \mathcal{E}_\ell} b_{\ell,e},
\end{equation}
where $b_{\ell,e} \ge 0$ is an additional propagation penalty encountered by propagation events $\mathcal{E}_\ell$ along the branch trajectory $\Gamma_\ell$.
The first term in \eqref{eqn:cost} represents the accumulated propagation length, while the second term accounts for additional propagation penalties by propagation mechanism with different boundary conditions. 
Accordingly, the computable reference eikonal surrogate is defined as the lower envelope of the propagation-mechanism-aware costs over all admissible branches
\begin{equation}\label{eqn:S_tilde}
    \tilde{S}_{\mathrm{ref}} (\bm x) = \min_{\ell \in \mathcal{A}(\bm x) }  J_\ell(\bm x),
\end{equation}
with the corresponding selected branch as
\begin{equation}\label{eqn:}
    \tilde{\ell}_{\mathrm{ref}} = \arg\min_{\ell \in\mathcal A(\bm x)} J_\ell(\bm x).
\end{equation}

\subsection{Eikonal Consistency Analysis and Critical Regions}\label{sec:critical}

The constructed reference eikonal function in \eqref{eqn:S_tilde} does not generally satisfy the eikonal relation globally, since spatial changes in the active minimizing branch or admissible branch set can introduce nonsmooth transitions. 
To establish its local physical consistency and characterize the critical structure, we first consider a smooth propagation region in which the selected branch $\ell$ remains unchanged.
Thus, within this region,
\begin{equation}
    \tilde S_{\mathrm{ref}}(\bm x)
    = J_\ell (\bm x) = 
    L_\ell(\bm x) + \sum_{e \in \mathcal{E}_\ell} b_{\ell,e}. 
\end{equation}
Assuming the propagation event penalties $b_{\ell,e}$ are spatially constant within the corresponding branch region, the local gradient of the constructed reference eikonal \eqref{eqn:S_tilde} is determined by the accumulated eikonal length in \eqref{eqn:L} and satisfies the local eikonal relation
\begin{equation}
    \begin{aligned}
    \nabla \tilde{S}_{\mathrm{ref}} (\bm x) = \nabla L_\ell =  \nabla S_{\mathrm{ref}} (\bm x),\\
    \|\nabla\tilde{S}_{\mathrm{ref}} (\bm x)\| = \| \nabla S_{\mathrm{ref}} (\bm x) \| = n(\bm x),
\end{aligned}
\qquad \bm x\in\Omega_{\mathrm{sm}}, 
\end{equation}
where $\Omega_{\mathrm{sm}} \subseteq \Omega$ denotes regions in which a single propagation branch is dominant and differentiable. Thus $\tilde{S}_{\mathrm{ref}}$ is a classical eikonal solution within each regular single-branch region and preserves the propagation direction of the selected branch, thereby ensuring that the spatial-frequency suppression characterized in Proposition~\ref{prop1} holds locally.

Branch switching occurs when two or more admissible branches attain the same minimum value, thereby preserving the continuity of the lower envelope while introducing a discontinuity in its gradient. Specifically, at a branch-switching interface, multiple candidate branches simultaneously attain the common minimum cost in \eqref{eqn:S_tilde}, i.e., $J_\ell (\bm x) = J_i (\bm x) = \ldots = \tilde{S}_{\mathrm{ref}} (\bm x)$.
In such regions, $\tilde{S}_{\mathrm{ref}}$ is continuous but non-differentiable, i.e.,
\begin{equation}
    \nabla J_l (\bm x) \neq  \nabla J_i (\bm x), \qquad x \in \Sigma_{\mathrm{kink}},
\end{equation}
where $\Sigma_{\mathrm{kink}} \subseteq \Omega$ denotes the branch switching region in $\Omega$. Although $\tilde{S}_{\mathrm{ref}}$ remains continuous within $\Sigma_{\mathrm{kink}}$, the classical gradient of $\tilde{S}_{\mathrm{ref}}$ does not exist, and the eikonal equation is therefore not defined in such locations.

Discontinuities in the lower envelope arise when the active branch changes as the admissible branch set $\mathcal{A}(\bm x)$ itself changes across a propagation interface. Specifically, consider an interface separating two regions with admissible branch sets $\mathcal{A}^-(\bm x)$ and $\mathcal{A}^+(\bm x)$. When the active minimizing branch on one side becomes inadmissible across the interface, the branch selected from the admissible set on the other side could attain a different value. Consequently, the lower envelopes on the two side do not coincide at the interface, i.e.,
\begin{equation}
    \min_{\ell \in \mathcal{A}^+(\bm x)}{ J_{\ell} (\bm x) } \neq \min_{\ell \in \mathcal{A}^-(\bm x)}{ J_{\ell} (\bm x) }, 
    \qquad \bm x \in \Sigma_{\mathrm{jump}},
\end{equation}
where $\Sigma_{\mathrm{jump}}$ denotes the value jump region. An admissibility-induced mismatch across the interface can produce a discontinuity in $\tilde{S}_{\mathrm{ref}}$, where the classical spatial gradient is undefined.



\begin{figure}[t]
    \centering
    \includegraphics[width=0.58\linewidth]{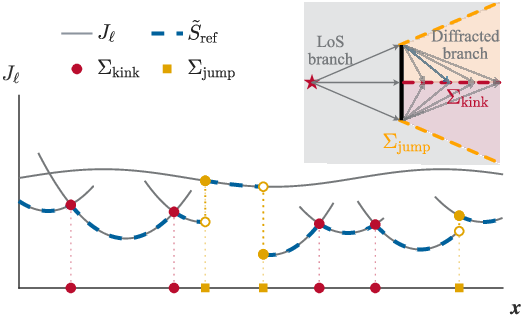}
    \caption{Illustration of the computable reference eikonal $\tilde{S}_\mathrm{ref}$, constructed as the lower envelope of branch-wise propagation-cost candidates $J_\ell$, with continuous branch-switching interfaces \(\Sigma_{\rm kink}\) and admissibility-induced jump interfaces \(\Sigma_{\rm jump}\).}
    \label{fig:lower_envelop}
\end{figure}

Overall, the constructed $\tilde{S}_{\mathrm{ref}}$ is a piecewise eikonal-consistent reference function, as illustrated in Fig~\ref{fig:lower_envelop}. Specifically, it satisfies the classical eikonal equation in the smooth region $\Omega_{\mathrm{sm}}$, is continuous but generally non-differentiable on $\Sigma_{\mathrm{kink}}$, and is discontinuous on $\Sigma_{\mathrm{jump}}$.
Consequently, the classical eikonal equation is applicable within regular single-branch regions $\Omega_{\mathrm{sm}}$, whereas the critical set $\Omega_\mathrm{crit} = \Omega \setminus \Omega_{\mathrm{sm}}$, comprising physical boundaries $\partial\Omega$, source location $\bm x_\mathrm{tx}$, continuous but non-differentiable switching interfaces $\Sigma_{\mathrm{kink}}$, and discontinuous admissibility transitions $\Sigma_{\mathrm{jump}}$ requires additional constraints.

\section{Propagation-Structure-Aware Radio Map Construction Framework}
\begin{figure*}[t]
    \centering
    \includegraphics[width=0.58\linewidth]{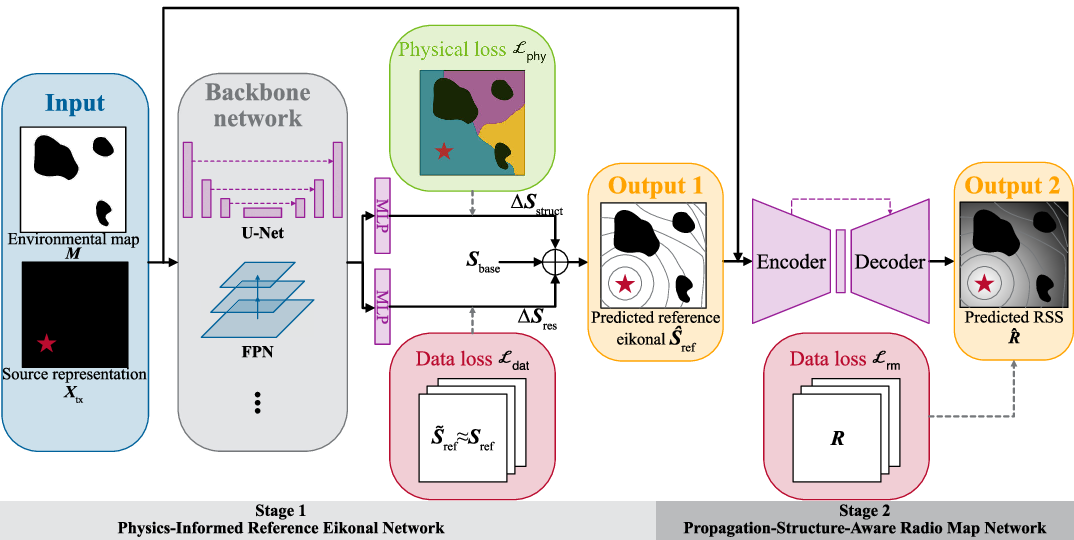}
    \caption{Architecture of the proposed two-stage framework. }
    \label{fig:NN_all}
\end{figure*}

Building on the reference eikonal construction and regularity analysis in Section~\ref{sec:III}, we develop a two-stage radio-map construction framework, as illustrated in Fig.~\ref{fig:NN_all}.
The first stage employs a physics-informed neural network to predict the reference eikonal under joint physics and data supervision, including scaffold constraints at critical regions.
The second stage exploits the resulting reference eikonal estimate as propagation-aware auxiliary information into a radio map construction network to predict the RSS distribution.


\subsection{Physics-Informed Reference Eikonal Network}

\subsubsection{Dual-Head Architecture with Structural–Residual Representation}
The first-stage network takes as input a multi-channel spatial tensor comprising the environmental map $\bm M$ and the transmitter representation $\bm X_{\mathrm{tx}}$, and predicts the sampled reference eikonal $\hat{\bm S}_{\mathrm{ref}}$, as illustrated in Fig.~\ref{fig:NN_all}. 
The network outputs a single-channel reference eikonal with the same spatial resolution as the input, expressed as
\begin{equation}
    \hat{\bm S}_{\mathrm{ref}} = \mathcal{N}_{\bm \vartheta}^\mathrm{eik} \left( \bm M, \bm X_{\mathrm{tx}} \right),
\end{equation}
where $\bm \vartheta$ denotes the parameters of the first-stage network. 


We apply a U-Net-based architecture comprising a shared encoder and two decoder heads.
The encoder extracts features at progressively coarser spatial resolutions, while dilated convolutions enlarge the receptive field to capture propagation structures extending across the environment. The decoder restores the original spatial resolution and incorporates encoder features through skip connections, after which two prediction heads produce the structural and residual outputs, respectively.

The reference eikonal characterizes the accumulated optical path associated with the selected propagation branch, whose spatial variation is determined by both the transmitter location and the surrounding propagation environment. 
In homogeneous free space with $n=1$, the eikonal of direct propagation from a point source is given by the Euclidean distance to the transmitter.
Motivated by this analytical structure, we introduce a simple geometric baseline to account for the source-distance dependence, allowing the network to focus on learning environment-dependent propagation corrections.
Accordingly, we define the Euclidean-distance baseline as
\begin{equation}
    S_{\mathrm{base}} (\bm x) = \|\bm x - \bm x_{\mathrm{tx}}\|.
\end{equation}
Building upon this baseline, the structure decoder is designed to capture the structural correction prescribed by the known physical model, whereas the residual decoder accounts for environment-dependent discrepancies that are not represented by the structural component.
Therefore, the predicted reference eikonal function is constructed by combining the baseline with the outputs of the two decoder heads
\begin{equation}
\begin{split}
    \hat{S}_{\mathrm{ref}} (\bm x) 
    &= \hat{S}_{\mathrm{struct}} (\bm x) + \Delta S_{\mathrm{res}} (\bm x) \\
    &= 
    S_{\mathrm{base}} (\bm x) + \Delta S_{\mathrm{struct}} (\bm x) + \Delta S_{\mathrm{res}} (\bm x),
\end{split}
\end{equation}
where $\Delta S_{\mathrm{struct}}$ and $\Delta S_{\mathrm{res}}$ denote the physics-consistent structural correction and environment-dependent residual correction, respectively, and $\hat{S}_{\mathrm{struct}} = S_{\mathrm{base}} (\bm x) + \Delta S_{\mathrm{struct}} (\bm x) $ is the predicted structured eikonal.

\subsubsection{Scaffold Training Strategy for Critical Regions}
As established in Section~\ref{sec:critical}, the constructed reference eikonal satisfies the classical eikonal equation within regular single-branch regions, while physical boundaries and other critical interfaces require separate treatment due to branch switching and admissibility transitions. 
Consequently, applying the classical eikonal residual over the entire domain would impose derivative constraints at locations where a single smooth propagation branch is not well defined, thereby introducing physically inconsistent supervision during training.

\begin{figure}[t]
    \centering
    \includegraphics[width=0.58\linewidth]{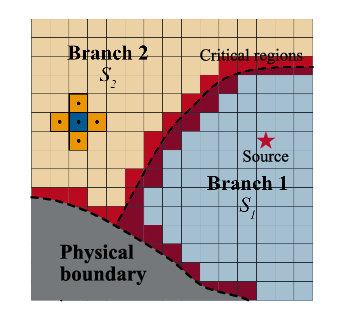}
    \caption{Scaffold training strategy for critical regions.}
    \label{fig:scaffold}
\end{figure}

As shown in Fig.~\ref{fig:scaffold}, to retain reliable physical guidance while avoiding invalid local derivative constraints, we introduce a scaffold training strategy, in which reference values at critical regions are used as anchor points. 
Let $\Omega^h$, $\Omega_{\mathrm{sm}}^h$, and $\Omega_{\mathrm{crit}}^h$ denote the discrete counterparts of the continuous spatial domains $\Omega$, $\Omega_{\mathrm{sm}}$, and $\Omega_{\mathrm{crit}}$, respectively, where $h$ denotes the spatial discretization resolution.
During training, the scaffold field is constructed as
\begin{equation}
    \hat{S}_{\mathrm{scaf}} (\bm x) = B_\mathrm{c} (\bm x) \tilde{S}_{\mathrm{ref}} (\bm x) + \left( 1 - B_\mathrm{c} (\bm x) \right) \hat{S}_{\mathrm{struct}} (\bm x) ,
\end{equation}
where
\begin{equation}
    B_\mathrm{c} (\bm x) = 
    \begin{cases}
        0, \qquad \bm x \in \Omega^h_\mathrm{sm}\\
        1, \qquad \bm x \in \Omega^h_\mathrm{crit}
    \end{cases}
\end{equation}
denotes the binary mask for critical regions. 
These anchors directly impose reference values at physically critical locations, while the classical eikonal residual is applied only within regular single-branch regions. 
As a result, the derivative within regular single-branch regions could be approximated using the central differences method,
\begin{equation}
\begin{split}
    \nabla_d \hat{S}_\mathrm{scaf} (\bm x) = \frac{\hat{S}_\mathrm{scaf} (\bm x + h \bm q_d) - \hat{S}_\mathrm{scaf} (\bm x - h \bm q_d)}{2h}, \, \bm x \in \Omega^h_\mathrm{sm},
\end{split}
\end{equation}
where $\bm q_d$ is the unit vector along the $d$-th coordinate direction. 

\subsubsection{Physics-Informed Training Objective}
The regional characterization in Section~\ref{sec:critical} motivates distinct physical constraints for smooth and critical propagation regions.
In smooth single-branch regions, the reference eikonal is locally differentiable and satisfies the classical eikonal equation, whereas this differential relation is not pointwise valid in critical regions due to boundaries, branch switching, or admissibility transitions.
Accordingly, we enforce eikonal consistency in smooth regions and use the constructed reference eikonal as scaffold anchors in critical regions
\begin{equation}
\begin{split}
    \mathcal{L}_{\mathrm{phy}}  = \omega_{\mathrm{sm}} \frac{1}{|\Omega^h_{\mathrm{sm}}|}
    \sum_{\bm{x}\in\Omega^h_{\mathrm{sm}}}
    \left|
    \left\|\nabla \hat{S}_{\mathrm{scaf}}(\bm{x})\right\|
    -
    n(\bm{x})
    \right|^2 \\
    + \omega_{\mathrm{crit}} \frac{1}{|\Omega^h_{\mathrm{crit}}|}
    \sum_{\bm{x}\in\Omega^h_{\mathrm{crit}}}
    \left| \hat{S}_{\mathrm{scaf}}(\bm{x}) - \tilde{S}_{\mathrm{ref}}(\bm{x}) \right|^2
    .
\end{split}
\end{equation}
where \(\omega_{\mathrm{sm}}\) and \(\omega_{\mathrm{crit}}\) weight the smooth- and critical-region constraints, respectively.
These complementary regional constraints preserve the applicable local propagation physics while providing the boundary and critical interface information required to determine the branch-structured reference field.

The physics loss promotes eikonal consistency in smooth regions and enforces scaffold-based constraints at critical regions, providing physics-guided structural supervision. To further account for environment-dependent modeling discrepancies and improve the accuracy of the constructed reference field, data-consistency loss is defined as
\begin{equation}
    \mathcal{L}_{\mathrm{dat}} = \frac{1}{|\Omega^h|}
    \sum_{\bm{x}\in\Omega^h}
    \left|
    \hat{S}_{\mathrm{ref}}(\bm{x})
    -
    \tilde{S}_{\mathrm{ref}}(\bm{x})
    \right|^2.
\end{equation}
In addition to providing supervision at critical regions, the data-consistency loss enables the residual head to capture discrepancies that are not explained by the physics-constrained structure decoder, including approximation errors in the physical model and environment-dependent effects.

The overall training objective of the physics-informed reference eikonal network is formulated as 
\begin{equation}
    \mathcal{L}_{\mathrm{ref}}
=
\mu_{\mathrm{phy}}\mathcal{L}_{\mathrm{phy}}
+
\mu_{\mathrm{dat}}\mathcal{L}_{\mathrm{dat}},
\end{equation}
where $\mu_{\mathrm{phy}}$ and $\mu_{\mathrm{dat}}$ balance the contributions of the physics-consistency and data-consistency losses, respectively.
Together, these complementary objectives form a physically consistent yet data-adaptive representation of the reference eikonal field.


\subsection{Propagation-Structure-Aware Radio Map Network}
The reference eikonal network provides a propagation-aware representation of the dominant spatial structure determined by the source and surrounding environment. 
The second-stage network then incorporates the predicted reference eikonal as a structural prior to estimate the spatial RSS distribution.

As shown in Fig.~\ref{fig:NN_all}, the propagation-structure-aware radio-map network takes as input the environmental representation, source information, and reference eikonal predicted by the first-stage network. The resulting RSS map is expressed as
\begin{equation}
    \hat{\bm R} = \mathcal{N}_{\bm \varphi}^\mathrm{rm} \left( \bm M, \bm X_\mathrm{tx}, \hat{\bm S}_\mathrm{ref} \right),
\end{equation}
where $\bm \varphi$ denotes the trainable parameters of the second-stage network.
Consistent with Proposition~\ref{prop1}, the predicted reference eikonal provides a physically structured propagation prior that facilitates learning the complex environment-to-RSS mapping.


The second-stage network employs a single-decoder U-Net to predict the RSS field from the combined propagation, environmental, and transmitter information. It is trained by minimizing the pointwise error between the predicted and ground-truth RSS maps.
\begin{equation}
    \mathcal{L}_\mathrm{rm} = \frac{1}{|\Omega^h|}
    \sum_{\bm{x}\in\Omega^h}
    \left|
    R(\bm{x})
    -
    \hat{R}(\bm{x})
    \right|^2.
\end{equation}
Together, the propagation-structure auxiliary representation and RSS supervision enable the second-stage network to capture environment-dependent spatial RSS variations while preserving the dominant propagation structure.

\subsection{Training and Inference Procedures}
Algorithm~\ref{alg:framework} summarizes the two-stage training and inference procedures of the proposed framework.
Notably, reference-label construction and physics-based constraints are confined to offline training, while online radio-map construction requires only sequential inference through the two trained networks, thereby maintaining low computational complexity.

\begin{algorithm}[t]
\caption{Pipeline of the Proposed Framework}
\label{alg:framework}
\scalebox{0.73}{
\begin{minipage}{1\textwidth}
\begin{algorithmic}[1]
\Statex \textbf{Offline Training}
\Require Training set $\{(\bm M_i,\bm X_i, \bm R_i)\}_{i=1}^{N}$
\State Construct reference labels $\tilde{\bm S}_{\mathrm{ref}}$
and critical-region masks $\bm B_\mathrm{c}$
\State Train reference network $\mathcal{N}_{\bm \vartheta}^\mathrm{eik}$ by minimizing $\mathcal L_{\mathrm{ref}}$
\State Generate predicted reference maps $\hat S_{\mathrm{ref}}$ using $\mathcal{N}_{\bm \vartheta^*}^\mathrm{eik}$
\State Train RSS network $\mathcal{N}_{\bm \varphi}^\mathrm{rm}$ using $\hat{\bm S}_{\mathrm{ref}}$ by minimizing $\mathcal L_{\mathrm{rm}}$
\Statex \textbf{Online Inference}
\Require Environment $\bm M$, transmitter location $\bm X$
\State $\hat{\bm S}_{\mathrm{ref}}
\gets \mathcal{N}_{\bm \vartheta^*}^\mathrm{eik}(\bm M,\bm X)$
\State $\hat{\bm R}\gets \mathcal{N}_{\bm \varphi}^\mathrm{rm}(\bm M,\bm X,\hat{\bm S}_{\mathrm{ref}})$
\State \Return $\hat{\bm R}$
\end{algorithmic}
\end{minipage}
}
\end{algorithm}

\section{Numerical Results and Discussion}

\subsection{Experimental Setup}
We evaluate the proposed framework through numerical experiments on both reference eikonal construction and RSS radio-map prediction. The experiments are designed to evaluate the effectiveness of the proposed physics-informed training strategy in capturing the underlying propagation structure and its resulting improvement in radio-map construction accuracy.

\subsubsection{Dataset}
The proposed framework is evaluated on the RadioMapSeer \cite{yapar2022dataset}, a publicly available dataset for urban wireless propagation modeling. 
The dataset comprises $701$ two-dimensional (2D) urban maps, each containing $80$ transmitter locations and their corresponding simulated radio maps. The radio maps are generated using the dominant path model (DPM) \cite{Wahl2005DominantPath} and intelligent ray tracing (IRT) \cite{Rautiainen2002RayTracing}. For IRT, up to two ray and geometry interactions are considered, hereafter referred to as IRT2. Each radio map is densely sampled on a $256 \times 256$ receiver grid with a spatial resolution of $1$ meter. Further details on the dataset configuration and simulation parameters can be found in \cite{yapar2022dataset}.

In addition to the RSS radio maps, the reference eikonal maps are generated for each transmitter location. Specifically, the computable reference eikonal field $\tilde{S}_{\mathrm{ref}}$ is constructed over the spatial domain according to \eqref{eqn:S_tilde}, providing propagation-structure supervision for reference-eikonal learning. The corresponding critical regions are further identified and used to impose scaffold constraints during network training.
Moreover, we generate a smaller auxiliary dataset using Sionna RT \cite{10465179} to provide high-fidelity ray-tracing references for physical validation. For each selected environment and transmitter configuration, multipath propagation information is extracted to identify the strongest-energy propagation branch and construct the corresponding reference eikonal and directional characteristics. Owing to the substantially higher computational cost of ray tracing, this dataset is not used for large-scale network training, but serves exclusively for validation purposes.



\subsubsection{Evaluation Metrics}
To comprehensively evaluate the proposed framework, we consider metrics from four aspects, i.e., accuracy of the constructed radio map, fidelity of the constructed reference eikonal, physical validity of the directional representation, and computational efficiency.

For the final RSS radio-map construction, we adopt the NMSE as the performance metric for the proposed framework and the compared methods. Given the ground-truth radio map $\bm R$ and its estimate $\hat{\bm R}$, the NMSE is defined as 
\begin{equation}
    \mathrm{NMSE} = \frac{\|\hat{\bm R} - \bm R \|^2}{ \| \bm R\|^2},
\end{equation}
where a lower NMSE indicates higher prediction accuracy.


To evaluate the fidelity of the constructed reference eikonal, we separately examine its behavior around critical regions and within regular propagation regions.
Let \(\tilde{\Omega}^h_{\mathrm{crit}}\) denote the critical-region set identified from the proposed computable reference eikonal field in \eqref{eqn:S_tilde}.
Their spatial discrepancy is quantified by the symmetric mean boundary distance (MBD),
\begin{equation}
\begin{split}
    d_\mathrm{MBD} = \frac{1}{2} 
    \Biggl(
    \frac{1}{|\tilde{\Omega}^h_{\mathrm{crit}}|} \sum_{\bm x \in \hat{\Omega}^h_{\mathrm{crit}}} \min_{\bm y \in \Omega^h_\mathrm{crit}} \| \bm x - \bm y \| \\
    + \frac{1}{|\Omega^h_\mathrm{crit}|} \sum_{\bm y \in \Omega^h_\mathrm{crit}} \min_{\bm x \in \hat{\Omega}^h_{\mathrm{crit}}} \| \bm x - \bm y \|
    \Biggr),
\end{split}
\end{equation}
where a smaller $d_\mathrm{MBD}$ indicates better localization of the critical interfaces.
Within regular regions, the directional consistency is evaluated through the angular error between the proposed and Sionna-RT-derived reference-eikonal gradients,
\begin{equation}
    e_\theta\left(\bm x\right) = \arccos \left( \frac{ \nabla \tilde{S}_\mathrm{ref}^T (\bm x) \nabla S_\mathrm{ref} (\bm x) }{\|\nabla \tilde{S}_\mathrm{ref} (\bm x)\| \|\nabla S_\mathrm{ref} (\bm x)\| } \right), \qquad \bm x \in \Omega^h_\mathrm{sm},
\end{equation}
where $\nabla S_\mathrm{ref}$ is the reference-eikonal gradient obtained from the Sionna RT results. A smaller angular error indicates better preservation of the dominant propagation direction.

To examine whether the directional condition underlying Proposition~\ref{prop1} is representative of the considered high-frequency propagation environments, we further define the directional power concentration ratio. 
The power concentration within the spatial frequency suppression angular region is defined as
\begin{equation}
    \rho_{\pi/3} (\bm x) = \frac{\sum_{\ell: \theta_\ell\leq \pi/3} |A_\ell (\bm x)|^2 }{\sum_{\ell \in \mathcal{A}(\bm x)} |A_\ell (\bm x)|^2}.
\end{equation}
A larger $\rho_{\pi/3}$ indicates that a greater fraction of the local multipath power is concentrated within the spatial-frequency-reduction sector centered on the reference direction.

Finally, the computational efficiency of reference-eikonal construction is evaluated using the average runtime required to generate one reference field. The proposed method and Sionna RT are executed under the same hardware configuration and comparable propagation settings.

\subsection{Validation of the Reference Propagation Structure}
\subsubsection{Reference-Eikonal Fidelity and Construction Efficiency}
\begin{figure}[t]
\centering
    \begin{subfigure}[t]{\linewidth}
    \centering
    \includegraphics[width=0.58\textwidth]{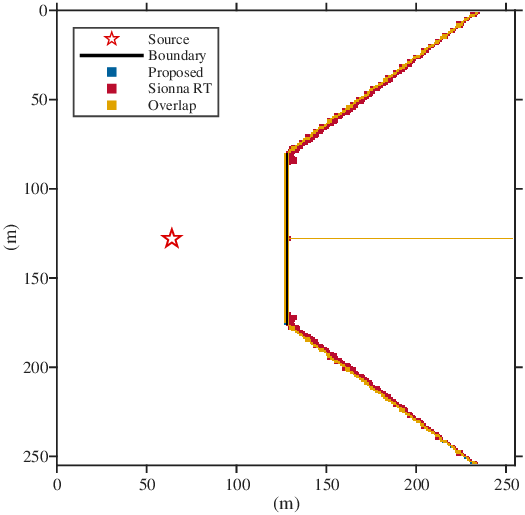}
    \caption{} \label{fig:fiedlity-a}
\end{subfigure}
\vfill
\begin{subfigure}[t]{\linewidth}
    \centering
    \includegraphics[width=0.58\textwidth]{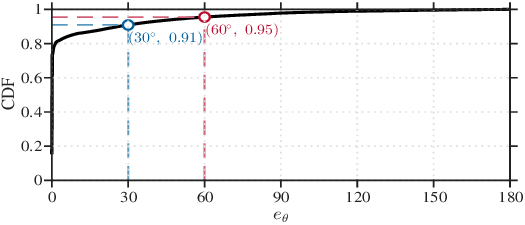}
    \caption{} \label{fig:fiedlity-b}
\end{subfigure}
\caption{Accuracy evaluation of the proposed computable reference eikonal against Sionna RT in critical and smooth regions: (a) spatial consistency of the identified critical regions in a representative environment; (b) empirical CDF of the gradient-direction error over regular regions.}
\label{fig:fiedlity}
\end{figure}

Fig.~\ref{fig:fiedlity} evaluates the accuracy of the computable reference eikonal against the Sionna RT reference.
In the representative example of Fig.~\ref{fig:fiedlity-a}, the critical regions identified by the proposed method closely match the Sionna-RT reference, including physical boundaries $\partial\Omega$, the source location $\bm{x}_{\mathrm{tx}}$, branch-switching interfaces $\Sigma_{\mathrm{kink}}$, and admissibility-transition interfaces $\Sigma_{\mathrm{jump}}$.
Additionally, quantitative evaluation across the dataset yields a median and mean MBD of 2.93 and 3.45 pixels, respectively.
In smooth regions, the cumulative distribution function (CDF) in Fig.~\ref{fig:fiedlity-b} shows gradient-direction errors below $30^\circ$ and $60^\circ$ at 91\% and 95\% of evaluated locations, respectively.
Meanwhile, the proposed construction requires only 6.99 s, compared with 2700 s for Sionna RT, achieving a 386-fold speedup under the evaluated configuration.
These results demonstrate that the proposed construction preserves both the critical region geometry and the dominant propagation direction with high fidelity while substantially reducing the computational cost.

\subsubsection{Directional Power Concentration}
\begin{figure}[t]
\centering
    \begin{subfigure}[t]{\linewidth}
    \centering
    \includegraphics[width=0.58\textwidth]{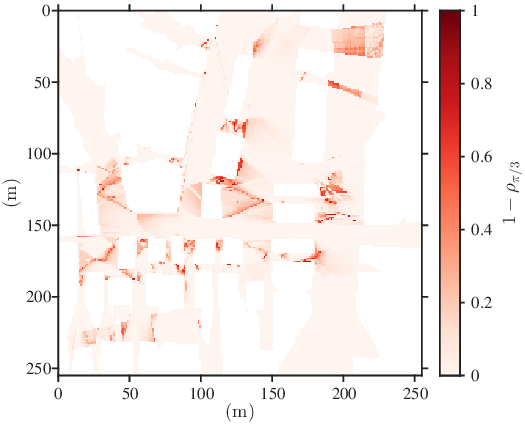}
    \caption{} \label{fig:red+energy_1}
\end{subfigure}
\vfill
\begin{subfigure}[t]{\linewidth}
    \centering
    \includegraphics[width=0.58\textwidth]{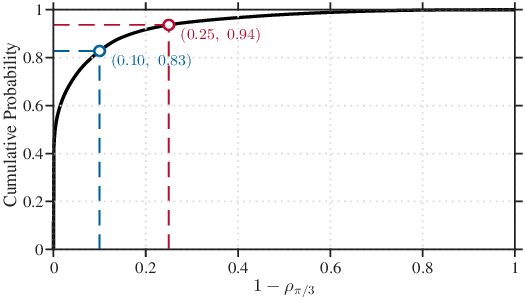}
    \caption{} \label{fig:red+energy_2}
\end{subfigure}
\caption{Directional power concentration around the reference eikonal branch under the IRT2 setting: (a) Spatial distribution of \(1-\rho_{\pi/3}\) in a representative scene. (b) Empirical CDF of \(1-\rho_{\pi/3}\) over the evaluated receiver locations.}
\label{fig:red+energy}
\end{figure}

Fig.~\ref{fig:red+energy} examines whether the directional condition underlying Proposition~\ref{prop1} is representative of the considered high-frequency propagation scenarios using Sionna RT simulations. 
The spatial distribution in Fig.~\ref{fig:red+energy_1} shows the multipath power is predominantly concentrated around the introduced reference eikonal direction, with a mean concentration ratio of $0.96$ and a median of $1.00$. 
This observation is further confirmed by the empirical CDF in Fig.~\ref{fig:red+energy_2}. Approximately $83\%$ of the evaluated locations concentrate at least $90\%$ of total multipath power within the $\pm \pi/3$ sector centered on the reference direction, and this proportion increases to $94\%$ when the threshold is relaxed to $75\%$. 
These results further justify selecting the strongest path as the reference, as most of the multipath power lies within the angular sector in which the reference-eikonal transformation reduces branch-wise spatial frequency.

\subsubsection{Sensitivity to Reference-Direction Mismatch}

\begin{figure}[t]
    \centering
    \includegraphics[width=0.58\linewidth]{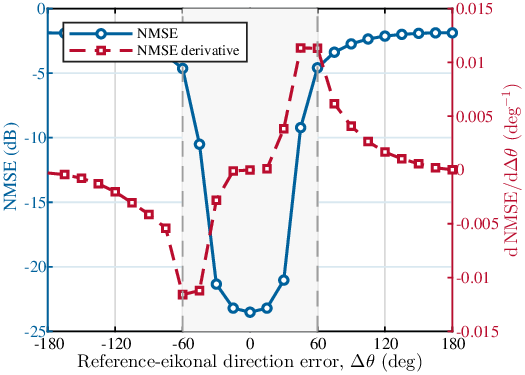}
    \caption{NMSE and its angular derivative versus the reference-eikonal direction offset $\Delta\theta$.}
    \label{fig:rotation}
\end{figure}

Fig.~\ref{fig:rotation} evaluates the sensitivity of radio-map construction to directional mismatch of the reference eikonal.
By introducing controlled angular offsets relative to the Sionna-RT reference direction, the NMSE remains nearly unchanged for small perturbations, confirming the robustness of the proposed method against moderate directional mismatch.
The error increases sharply as the mismatch approaches $\Delta \theta = \pm 60^\circ$, where the magnitude of the NMSE derivative is maximized.
This transition is qualitatively consistent with the angular condition in Proposition~\ref{prop1}, under which reference-eikonal transformation reduces the spatial frequency of a propagation branch.
These results therefore provide empirical support for both the robustness of the proposed reference representation and the directional mechanism underlying its performance gain.

\subsection{Radio-Map Construction Performance}
\subsubsection{Comparison with State-of-the-Art Methods}
Table~\ref{tab:performance_comparison} compares the proposed framework with representative radio-map construction methods under the DPM and IRT2 propagation settings. 
The proposed method consistently achieves the lowest NMSE in both settings, attaining 0.0026 on DPM and 0.0039 on IRT2. 
Compared with the strongest available baseline in each setting, namely RadioDiff-\(k^2\) on DPM and LPCGMN on IRT2, these results correspond to NMSE reductions of 39.53\% and 79.26\%, respectively. The consistent improvements across both settings support the effectiveness of the proposed framework for accurate radio-map construction.

\begin{table}[t]
\centering
\caption{NMSE Comparison on DPM and IRT2 and performance gains achieved by the proposed method.}
\label{tab:performance_comparison}
\scalebox{0.73}{
\begin{tabular}{lcccc}
\toprule
\multirow{2}{*}{Method}
& \multicolumn{2}{c}{DPM}
& \multicolumn{2}{c}{IRT2} \\
\cmidrule(lr){2-3} \cmidrule(lr){4-5}
& NMSE & Reduction & NMSE & Reduction \\
\midrule
RadioUNet \cite{9354041}
& 0.0075 & 65.33\%
& 0.0219 & 82.19\% \\

LPCGMN \cite{10791446}
& 0.0064 & 59.38\%
& 0.0188 & 79.26\% \\

RadioMamba \cite{11190042}
& 0.0050 & 48.00\%
& 0.0262 & 85.11\% \\

RadioDiff-$k^2$ \cite{11278649}
& 0.0043 & 39.53\%
& $^{*}$ & $^{*}$ \\
\textbf{Proposed}
& \textbf{0.0026} & --
& \textbf{0.0039} & -- \\
\bottomrule
\end{tabular}}
\begin{tablenotes}
\footnotesize
\item$^{*}$ Not reported and not reproduced due to the high cost of diffusion-model training.
\end{tablenotes}

\end{table}

\subsubsection{Transferability Across Backbones}
Table~\ref{tab:backbone_prior_comparison} evaluates the backbone dependence of the proposed reference-eikonal guidance. 
For this purpose, we employ several generic backbones implemented using standard module components, including U-Net, ResNet, ViT, and Mamba.
Incorporating the predicted reference eikonal consistently improves prediction accuracy across all evaluated backbones.
The relative NMSE reductions range from 41.54\% to 89.52\% under DPM and from 85.54\% to 93.11\% under IRT2. Nevertheless, every backbone benefits from the proposed reference eikonal, supporting its utility as a transferable auxiliary representation across network architectures. The consistently large gains under IRT2 further indicate its value in the evaluated multipath setting.

\begin{table}[t]
\centering
\caption{NMSE comparison across different radio map construction network backbones with and without the predicted reference eikonal on DPM and IRT2.}
\label{tab:backbone_prior_comparison}
\scalebox{0.73}{
\begin{tabular}{lcccccc}
\toprule
\multirow{2}{*}{Backbone}
& \multicolumn{3}{c}{DPM}
& \multicolumn{3}{c}{IRT2} \\
\cmidrule(lr){2-4} \cmidrule(lr){5-7}
& w/o $\hat{S}_\mathrm{ref}$
& w/ $\hat{S}_\mathrm{ref}$
& Reduction
& w/o $\hat{S}_\mathrm{ref}$
& w/ $\hat{S}_\mathrm{ref}$
& Reduction \\
\midrule
U-Net
& 0.042 & 0.0044 & 89.52\%
& 0.13  & 0.0090 & 93.08\% \\

ResNet
& 0.021 & 0.0067 & 68.10\%
& 0.079 & 0.0074 & 90.63\%\\

ViT
& 0.029 & 0.011 & 62.07\%
& 0.097 & 0.011 & 88.66\% \\

Mamba
& 0.013 & 0.0076 & 41.54\%
& 0.046 & 0.0066 & 85.65\% \\
\bottomrule
\end{tabular}
}
\end{table}

\subsubsection{Training-Data Efficiency}
\begin{figure}[t]
    \centering
    \includegraphics[width=0.58\linewidth]{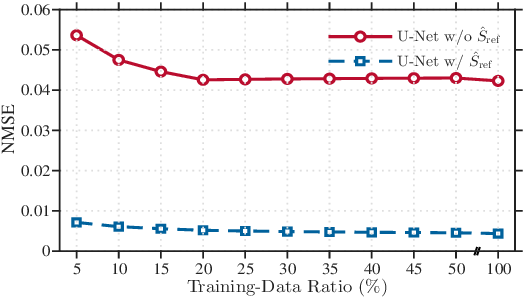}
    \caption{ NMSE versus the fraction of training data for U-Net with and without predicted reference-eikonal on DPM.}
    \label{fig:fig_dataScale}
\end{figure}

Fig.~\ref{fig:fig_dataScale} evaluates the data efficiency provided by the predicted reference eikonal.
The reference-eikonal input consistently improves construction accuracy across all evaluated training-data ratios.
With only 5\% of the training data, the proposed model outperforms the baseline without the reference-eikonal input trained on the full dataset.
This result demonstrates that the reference eikonal provides informative propagation-structure information that substantially reduces the amount of training data required for accurate RSS radio-map construction.

\subsection{Ablation Studies}

\subsubsection{Reference Eikonal Network Architecture and Training}
Table~\ref{tab:net1_ablation} evaluates the contributions of the major components to reference-eikonal learning.
Under physics-loss training, introducing scaffold anchors reduces NMSE from 0.0030 to 0.0027, corresponding to a 10.00\% reduction. This improvement supports the role of critical-region anchors in supplementing local eikonal constraints with reference-value information. 
Data-only training also achieves an NMSE of 0.0027, while the complete method further reduces it to 0.0026. This modest additional gain suggests that physics regularization and data supervision provide complementary guidance for learning the reference field.
The dual-head configuration also achieves a lower NMSE than the single-head configuration, reducing the error from 0.0036 to 0.0026, which supports separating the physics-regularized structural prediction from the residual correction.
Overall, the results demonstrate the complementary contributions of local eikonal consistency, critical-region scaffold constraints, data supervision, and the dual-head architecture to accurate reference-eikonal prediction.


\begin{table}[t]
\centering
\caption{Ablation study of the proposed physics-informed reference-eikonal network design.}
\label{tab:net1_ablation}
\scalebox{0.73}{
\begin{tabular}{lc}
\toprule
Configuration & NMSE \\
\midrule

Physics loss only, w/o scaffold
& 0.0030\\

Physics loss only, w/ scaffold
& 0.0027 \\

Data loss only
& 0.0027 \\

Single-head
& 0.0036 \\

\textbf{Proposed}
& \textbf{0.0026} \\

\bottomrule
\end{tabular}
}
\end{table}

\subsubsection{Effect of Auxiliary Propagation Representations}
Table~\ref{tab:input_ablation} evaluates the contribution of different auxiliary propagation representations to RSS radio-map construction. 
Introducing the Euclidean-distance representation $S_{\mathrm{base}}$ reduces the NMSE from 0.0071 to 0.0040, corresponding to a 43.66\% reduction, showing that coarse source-distance spatial information already provides useful geometric guidance for radio-map prediction.
Incorporating the gradient of the predicted reference eikonal, $\nabla \hat{S}{\mathrm{ref}}$, further reduces the NMSE to 0.0027, while the complete predicted reference field $\hat{S}{\mathrm{ref}}$ achieves an NMSE of 0.0026, corresponding to reductions of 61.97\% and 63.38\%, respectively, relative to using no auxiliary input.
The comparable performance of $\nabla \hat{S}{\mathrm{ref}}$ and $\hat{S}{\mathrm{ref}}$ indicates that local propagation-direction information constitutes a major source of the reference field's benefit, consistent with the directional spatial-frequency mechanism characterized in Proposition~\ref{prop1}.
Finally, using the computable reference eikonal $\tilde{S}{\mathrm{ref}}$ as an oracle input achieves an NMSE of 0.0025, only marginally lower than the 0.0026 achieved by $\hat{S}{\mathrm{ref}}$, indicating that the learned reference eikonal preserves most of the propagation information provided by the constructed field for downstream RSS construction.


\begin{table}[t]
\centering
\caption{Ablation study of auxiliary input representations for the radio-map prediction network under the DPM setting.}
\label{tab:input_ablation}
\scalebox{0.73}{
\begin{tabular}{ccc}
\toprule
Auxiliary Input & NMSE  & Reduction  \\
\midrule
None 
& $0.0071$
& -- \\

$S_{\mathrm{base}}$
& $0.0040$
& 43.66\% \\

$\nabla \hat{S}_{\mathrm{ref}}$
& $0.0027$
& 61.97\% \\

$\hat{S}_{\mathrm{ref}}$ (\textbf{Proposed})
& $0.0026$
& 63.38\% \\

$\tilde{S}_{\mathrm{ref}}$ (Oracle)
& $0.0025$
& 64.79\% \\
\bottomrule
\end{tabular}
}
\end{table}

\section{Conclusions}
We have developed a propagation-structure-aware framework for RSS radio-map construction in high-frequency multipath environments.
A reference eikonal was introduced to capture the dominant propagation structure, and its effectiveness was characterized through the relationship between directional mismatch and residual spatial frequency. 
An efficient computable reference eikonal was further constructed and integrated into a two-stage physics-informed construction framework with dedicated constraints on smooth and critical propagation regions. 
Numerical results demonstrated accurate and computationally efficient reference-structure construction, robustness to moderate directional mismatch, and consistent performance gains across different propagation settings and network backbones.
These results highlight that compact and physically meaningful propagation representations can be more effective than simply increasing model complexity or incorporating increasingly detailed propagation information.





{\appendix \label{append:proof}
By the law of cosines and the local eikonal relations $|\nabla S_\ell|=|\nabla S_{\mathrm{ref}}|=n$, the normalized residual spatial-frequency ratio is given by
\begin{equation}
    \gamma_\ell = 
    \frac{ \| \tilde{\bm \kappa}_\ell \|}{ \| \bm \kappa_\ell \|} = 
    \frac{\left\| \nabla S_\ell  - \nabla S_{\mathrm{ref}}  \right\|}{\left\| \nabla S_\ell  \right\|} =
    \sqrt{2 \left( 1 - \cos{\theta_\ell} \right)},
\end{equation}
which is equivalent to \eqref{eqn:gamma}.
For every fixed $m\geq1$, repeated application of the product rule gives
\begin{equation}
    \| \nabla^m \tilde{u}_\ell (\bm x) \| \sim |A_\ell (\bm x)|  \left( \gamma_\ell \| \bm \kappa_\ell  (\bm x) \| \right)^m,
\end{equation}
which establishes \eqref{eqn:tilde_u_l'm}.
To characterize the condition under which the transformation suppresses the spatial oscillation of the $\ell$-th branch, note that, since $m\geq1$ and $\gamma_\ell\geq0$, the leading $m$-th-order derivative magnitude is reduced if and only if 
\begin{equation}
    \gamma_\ell < 1.
\end{equation}
Using \eqref{eqn:gamma}, this condition is equivalent to
\begin{equation}
    \sin\left(\frac{\theta_\ell}{2}\right)<\frac{1}{2}.
\end{equation}
Since $\sin(\theta_\ell/2)$ increases monotonically on $\theta_\ell \in [0,\pi]$, the inequality yields \eqref{eqn:theta}.
Hence, the reference eikonal transformation strictly reduces the high-frequency spatial variation of the $\ell$-th branch if and only if its propagation direction deviates from the reference direction by less than $60^\circ$.}



 
%

\bibliographystyle{IEEEtran}
\bibliography{Ref}

 




\vfill

\end{document}